\documentclass[
preprint,
showpacs,
preprintnumbers,
tightenlines,
 amsmath,amssymb,
 aps]{revtex4-1}

\usepackage{hyperref}
\usepackage{graphicx}
\usepackage{url}

\newcommand{\OO}[1]{{\mathcal O}(c^{-#1})}

\newcommand{\ve}[1]{\mbox{\boldmath$#1$}}

\let\fl\relax
\let\ead\email

\def\exter{+}
\def\inter{-}

\let\oldbibitem\bibitem
\renewcommand\bibitem[2][]{\oldbibitem{#2}}

\def\bdis{\begin{displaymath}}
\def\edis{\end{displaymath}}

\def\beq{\begin{equation}}
\def\eeq{\end{equation}}

\def\nhat{{\hat n}}

\def\ve#1{{\bf #1}}
\def\OO#1{{\cal O}(c^{-#1})}

\begin{document}

\title[The post-linear Schwarzschild solution in harmonic coordinates]
{The post-linear Schwarzschild solution in harmonic coordinates:\\ elimination
of structure-dependent terms}

\author{Sergei A. Klioner}

\ead{Sergei.Klioner@tu-dresden.de}

\author{Michael Soffel}

\address{Lohrmann Observatory, Technische Universit\"at Dresden,
Mommsenstra\ss{}e 13, 01062 Dresden, Germany}

\begin{abstract}
The paper deals with a special kind of problems that appear in
solutions of Einstein's field equations for extended bodies: many
structure-dependent terms appear in intermediate calculations that
cancel exactly in virtue of the local equations of motion or can be
eliminated by appropriate gauge transformations. For a single body at
rest these problems are well understood for both the post-Newtonian
and the post-Minkowskian cases. However, the situation is still
unclear for approximations of higher orders. This paper discusses this
problem for a ``body'' of spherical symmetry to post-linear order.  We
explicitly demonstrate how the usual Schwarzschild field can be
derived directly from the field equations in the post-linear
approximation in the harmonic gauge and for an arbitrary spherically 
symmetric matter distribution. Both external and internal solutions
are considered.  The case of static incompressible fluid is then
compared to the well-known results from the literature.  The results
of this paper can be applied to generalize the well-known
post-Newtonian and post-Minkowskian multipole expansions of the metric
in the post-linear approximation.
\end{abstract}

\pacs{04.20.Jb, 04.25.Nx, 04.80.Cc, 95.10.Jk}

\maketitle

\tableofcontents

\newpage

 \section{Introduction}

There might be several reasons for an interest in the post-linear
Schwarzschild problem. Our main interest for that comes from the
problem of high-accuracy astrometry in the framework of General Relativity. 
Recently a series of high-accuracy astrometric space missions were proposed such as Gaia
\cite{PerrymanEtAl2001} with accuracies of a few microarcsecond
($\mu$as) or the Nearby Earth Astrometric Telescope (NEAT) proposed to
ESA \cite{MalbetEtAl2012}, for which accuracies around 50 nanoarcseconds
(nas) are under discussion. For all of these missions the
light propagation has to be calculated at a very high level of
accuracy that lies beyond the level of 1 $\mu$as in observed
directions.  Already for a mission like Gaia the influence of the
oblateness (quadrupole moment) of the bodies as well as their
barycentric motion cannot be neglected. Largest post-post-Newtonian
effects in the light propagation also have to be taken into account
\cite{KlionerZschocke2010}. Astrometric missions with angular
accuracies beyond 1 $\mu$as will certainly come in the near future and
also the day, when the subtle effects of higher post-Newtonian level
will be required. For those reasons it is of great importance to have
a metric tensor for a system of $N$ gravitationally interacting
arbitrarily shaped and composed, deformable and rotating bodies to
second post-Newtonian or second post-Minkowskian order (keeping all
terms in the velocities but only linear and quadratic terms in the gravitational
constant). Such a metric will form the basis for the modeling of light
trajectories. Some first steps toward such a metric have been done
\cite{XuWu2003,MinazoliChauvineau2009}) but the problem is far from
being solved. Clearly, further work is needed.

Tremendous work in General Relativity has been done with the harmonic
gauge that was found to be a useful and simplifying gauge for many
kinds of applications. It is logical to continue using the harmonic gauge
for further refinements of the theory needed for the high-accuracy astrometry
and celestial mechanics. The harmonic condition is defined by the following
equation ($g={\rm det}(g_{\alpha\beta})$ is the determinant of the metric tensor $g_{\alpha\beta}$):
\begin{eqnarray}\label{harmonic-gauge}
{\partial\over \partial x^\alpha}\left( {(-g)}^{1/2} g^{\alpha\beta} \right)=0\,.
\end{eqnarray}
\noindent
Several equivalent forms of the harmonic conditions can be found e.g. 
in Section 7.4 of \cite{Weinberg1972}.

For some 'body' (which in principle can be composed of a whole set of
individual 'bodies') at rest the external metric in the harmonic gauge
that is fully specified by two families of multipole moments, mass and
spin moments ($M_L$ and $S_L$) is known for both, the post-Newtonian
\cite{BlanchetDamour1989} and the post-Minkowskian cases
\cite{DamourIyer1991}.  For a system of point-like masses the whole
post-Minkowskian problem, the metric in harmonic coordinates as well
as the light-ray trajectories was solved in
\cite{KopeikinSchaefer1999}. This work was extended by including the
spin-monopoles of the bodies by Kopeikin \& Mashhoon
\cite{KopeikinMashhoon2002}. Kopeikin et
al. \cite{KopeikinKorobkovPolnarev2006} found an analytical
post-Minkowskian solution for the light propagation in the field of an
extended body {\it at rest}; here the full multipole structure was
taken into account.

Problems arise that are related with the internal structure of the bodies. For a
single body at rest these problems are well understood for both the
post-Newtonian and the post-Minkowskian cases \cite{BlanchetDamour1989,DamourIyer1991}
in which many structure dependent terms appear in
intermediate calculations that cancel exactly in virtue of the local
equations of motion or can be eliminated by corresponding gauge
transformations. However, for the post-linear case the situation is still
unclear. In the course of our studies for the general problem mentioned above we
found that even for the spherically symmetric case of a single body the
complete derivation of the external metric (the Schwarzschild metric) is
interesting.

We use fairly standard notations: $G$ is the Newtonian constant of
gravitation, $c$ is the vacuum speed of light. We use the signature
$(-+++)$ throughout this paper. Lower-case Latin indices
$i$, $j$, \dots take values 1, 2, 3. Lower-case Greek indices $\mu$,
$\nu$, \dots take values 0, 1, 2, 3. Repeated indices imply the
Einstein's summation irrespective of their positions
(e.g., $a^i\,b^i=a^1\,b^1+a^2\,b^2+a^3\,b^3$ and
$a^\alpha\,b^\alpha=a^0\,b^0+a^1\,b^1+a^2\,b^2+a^3\,b^3$).  We use
two special objects: $\delta^{ij}={\rm diag}(1,1,1)$ is the
Kronecker delta, $\varepsilon_{ijk}$ is the fully antisymmetric
Levi-Civita symbol ($\varepsilon_{123}=+1$). 
The 3-dimensional coordinate quantities (``3-vectors'') referred to
the spatial axes of the corresponding reference system are set in
boldface: $\ve{a}=a^i$.  
The scalar product of any two ``3-vectors'' $\ve{a}$ and $\ve{b}$
with respect to the Euclidean metric $\delta_{ij}$ is denoted by
$\ve{a}\,\cdot\,\ve{b}$ and can be computed as
$\ve{a}\,\cdot\,\ve{b}=\delta_{ij}\,a^i\,b^j=a^i\,b^i$.
A comma before an index designates the partial derivative with respect
to the corresponding coordinates: $A_{,\mu}={\partial
  A(t,\ve{x})/\partial x^\mu}$, $A_{,i}={\partial A(t,\ve{x})/\partial
  x^i}$. For partial derivatives with respect to the coordinate times
$t$ we use $A_{,t}={\partial A(t,\ve{x})/\partial t}$. A dot over any
quantity designates the total derivative with respect to the
coordinate time of the corresponding reference system: e.g. $\dot
A=dA/dt$.
Parentheses surrounding a group of indices denote symmetrization,
e.g., $A_{(ij)}={1\over 2}\left(A_{ij}+A_{ji}\right)$. Angle brackets
surrounding a group of indices or, alternatively, a caret on top of a
tensor symbol denote the symmetric trace-free (STF) part of the
corresponding object, e.g., $\hat A_{ij} \equiv A_{\langle ij
  \rangle}\equiv STF_{ij} A_{ij}= A_{(ij)}-{1\over
  3}\,\delta^{ij}\,A_{kk}$. For sequences of spatial indices we shall
use multi-indices: a spatial multi-index containing $l$ indices is
denoted by the same Latin character in the upper case $L$ ($K$ for $k$
indices, etc.): $L=i_1\dots i_l$, where each Cartesian index takes
values 1, 2, 3. We use also $L-1=i_1\dots i_{l-1}$, etc. A
multi-summation is understood for repeated multi-indices: $A_L\,B_L
\equiv \sum_{i_1\dots i_l} A_{i_1\dots i_l}\,B_{i_1\dots i_l}$. For a
spatial vector $v^i$ we denote $v^L\equiv v^{i_1}\,v^{i_2}\dots
v^{i_l}$. For an $L$-order partial derivative we denote
$\partial_L\equiv \partial_{i_1}\dots\partial_{i_l}$. For true
tensorial quantities like the energy-momentum tensor $T^{\mu\nu}$ or the
metric tensor $g_{\mu\nu}$ the position of each index (spatial or not)
is of great importance. For certain other quantities, like e.g. $w^i$,
$\sigma^{ij}$, or $q^{ij}$ introduced below, the position of indices
(upper or lower) is irrelevant (e.g., $w_i=w^i$).

In Section~\ref{section-general} we deal with the most generic case of an
arbitrary spherically symmetric mass distribution. The special case
of a static spherically symmetric incompressible matter distribution
will be treated in Section~\ref{section-incompressible-metric}
and the results will be compared to those known from the literature in
Section~\ref{section-incompressible}. Conclusions are formulated in 
Section~\ref{section-conclusions}.

\section{The general spherically symmetric case}
\label{section-general}

In this Section we deal with the most general case. Our goal is to
derive the external metric in the post-linear approximation for a
general spherically symmetric compact matter distribution.  From
Birkhoff's theorem it is clear that this external metric will be the
usual Schwarzschild metric that is determined by a single parameter,
the mass of the central body. The central point of this paper is to
demonstrate how other terms related with the structure of the body
(e.g., its radius $R$) that appear in intermediate calculations cancel
exactly or can be removed by a suitable gauge transformation. Other
aspects of the problem related with the usage of harmonic coordinates
are also of general interest.

\subsection{Metric tensor and field equations}

The post-linear metric tensor in harmonic coordinates will be written in the
form
\begin{eqnarray}
\label{g00}
g_{00} &=& -1 + {2 \over c^2} w - {2 \over c^4} w^2 + \OO6 \, ,
\\
\label{g0i}
g_{0i} &=& - {4 \over c^3} w^i + \OO5 \, , \\
\label{gij}
g_{ij} &=& \delta_{ij} \left( 1 + {2 \over c^2} w + {2 \over c^4} w^2 \right)
+ {4 \over c^4} q_{ij} + \OO5 \, .
\end{eqnarray}
\noindent
Here, the metric potentials $w$, $w^i$ and $q_{ij}$ obey the equations (see, e.g.,
\cite{MinazoliChauvineau2009})
\begin{eqnarray}
\label{wfieldeq}
\label{feq-w}
&&\Delta w - {1 \over c^2} w_{,tt} = - 4 \pi G \sigma + \OO4 \, ,
\\
\label{feq-wi}
&&\Delta w^i = - 4 \pi G \sigma^i + \OO2 \, , \\
\label{feq-qij}
\label{qij}
&&\Delta q_{ij} =  - w_{,i} w_{,j} - 4 \pi G \sigma^{ij} + \OO1 \, ,
\end{eqnarray}
where
\begin{equation}
\sigma = {T^{00} + T^{ss} \over c^2} \, , \quad \sigma^i = {T^{0i}
\over c} \, , \quad \sigma^{ij} = T^{ij} - \delta_{ij}\,T^{ss}
\end{equation}
\noindent 
and $T^{\mu\nu}$ are the components of the energy-momentum tensor.
The metric potentials $w$ and $w^i$ in (\ref{g00})--(\ref{gij})
are needed to orders $\OO2$ and ${\cal O}(c^{0})$, respectively.

\subsection{Formal solution of the field equations}

We consider an isolated compact matter distribution and, as usual, require
space-time to be asymptotically flat and covered by one single global
coordinate system $x^\mu = (ct, x^i)$ with
\begin{equation}\label{BRS-limits}
\lim_{|\ve{x}|\rightarrow \infty \atop{t = {\rm const.}}} g_{\mu\nu} = \eta_{\mu\nu}\,,
\end{equation}
\noindent
where $\eta_{\mu\nu}={\rm diag}(-1,+1,+1,+1)$ is the flat metric tensor
of Minkowski space-time. For this reason the field equations
should be solved with the boundary conditions
\begin{eqnarray}\label{w-mu-limits}
\lim_{|\ve{x}|\rightarrow \infty \atop{t = {\rm const.}}}
w(t,\ve{x}) = 0\,, \quad
\lim_{|\ve{x}|\rightarrow \infty \atop{t = {\rm const.}}}
w^i(t,\ve{x}) = 0\,, \quad
\lim_{|\ve{x}|\rightarrow \infty \atop{t = {\rm const.}}}
q_{ij}(t,\ve{x}) = 0\,.
\end{eqnarray}
\noindent 
The solution of (\ref{feq-w})--(\ref{feq-qij}) 
satisfying these boundary conditions that will be used in the following reads:
\begin{eqnarray}
\label{w-solution} 
\fl 
w(t,\ve{x})&=& G \int_V {\sigma(t,\ve{x}^\prime) \over
|\ve{x}-\ve{x}^\prime|}d^3x^\prime + {1\over 2 c^2}\,G\,{\partial^2 \over
\partial t^2}\, \int_V \sigma(t,\ve{x}^\prime) | \ve{x} - \ve{x}^\prime |
d^3x^\prime+\OO4\,,
\\
\label{wi-solution}
\fl
w^i(t,\ve{x})&=& G \int_V {\sigma^i(t,\ve{x}') \over |\ve{x}-\ve{x}'|}d^3x^\prime+\OO2\,,
\\
\label{qij-solution}
\fl
q_{ij}(t,\ve{x})&=&
{1\over 4\pi} \int_V {w_{,i}(t,\ve{x}')\,w_{,j}(t,\ve{x}') \over |\ve{x}-\ve{x}'|}d^3x^\prime
+G \int_V {\sigma^{ij}(t,\ve{x}') \over |\ve{x}-\ve{x}'|}d^3x^\prime+\OO1\,.
\end{eqnarray}
\noindent
Here $V$ is the support of the matter distribution.

\subsection{Spherically symmetric compact matter distribution}
\label{section-matter}

In the following we consider a matter distribution for which $T^{\mu\nu}$ has
compact support, that is in our reference system $(t,\ve{x})$ there exists
a quantity $R>0$ so that for $r\equiv|\ve{x}|>R$ the energy-momentum tensor
vanishes, $T^{\mu\nu}(t,\ve{x})=0$. In the following the matter located within
the area $|\ve{x}|\le R$ will be often called 'body'. Moreover, we will
consider a spherically symmetric matter distribution for which at an
arbitrary point one has
\begin{eqnarray}
\label{sigma}
\sigma&=&{T^{00} + T^{ss} \over c^2}=\sigma(t,r)\,
\\
\label{sigma-i}
\sigma^i&=&{1\over c}\,T^{0i}=B(t,r)\,n^i,\quad B(t,r)={1\over
c}\,T^{0i}\,n^i\,,
\\
\label{T-ij}
T^{ij}&=&A(t,r)\,\hat{n}^{ij}+\delta^{ij}\,C(t,r)\,,\quad
A(t,r)={3\over 2}T^{ij}\hat{n}^{ij},\ C(t,r)={1\over 3}\,T^{kk}\,,
\\
\label{sigma-ij}
\sigma^{ij}&=&A(t,r)\,\hat{n}^{ij}-2\,\delta^{ij}\,C(t,r)\,.
\end{eqnarray}
\noindent 
This form of the energy-momentum tensor is in agreement with the most
general form of the spherically symmetric metric tensor (see e.g.,
Section 13.5 of \cite{Weinberg1972}) and the corresponding field
equations. Thus, matter is fully characterized by 4 independent
scalar functions of time $t$ and radial coordinate $r=|\ve{x}|$:
$\sigma(t,r)$, $A(t,r)$, $B(t,r)$, and $C(t,r)$. No further
assumptions on these four functions are made. The body might be
non-static, it can oscillate or collapse etc. In the calculations
below the time $t$ plays a role as an additional parameter and we will
often omit the explicit dependence of these functions on time.

\subsection{Computation of the gravitational potentials $w$ and $w^i$}

The gravitational potentials $w$ and $w^i$ in the required approximation have
been extensively discussed in the literature. Here we summarize the results
needed for our further work. We specialize Eqs.\
(\ref{w-solution})--(\ref{wi-solution}) for the case of the spherically
symmetric matter distribution from Section \ref{section-matter}:
\begin{eqnarray}\label{w-solution-spherical}
\fl
w(t,\ve{x})&=& G\,r^2 \int_0^{2\pi}d\lambda^\prime\int_0^{\pi}d\theta^\prime\,
\sin\theta^\prime\,\int_0^{R/r}dz\,{z^2\, \sigma(t,z\,r)\over
\sqrt{1+z^2-2z\,\ve{n}^\prime\cdot\ve{n}}}
\nonumber\\
\fl
&&
+ {1\over 2 c^2}\,G\,r^4\,{\partial^2 \over \partial t^2}\,
\int_0^{2\pi}d\lambda^\prime\int_0^{\pi}d\theta^\prime\,
\sin\theta^\prime\,\int_0^{R/r}dz\,z^2\sigma(t,z\,r)\,
\sqrt{1+z^2-2z\,\ve{n}^\prime\cdot\ve{n}}
\nonumber\\
\fl
&&
+\OO4\,,
\\
\label{wi-solution-spherical}
\fl
w^i(t,\ve{x})&=& G\,r^2 \int_0^{2\pi}d\lambda^\prime\int_0^{\pi}d\theta^\prime\,
\sin\theta^\prime\,n^{\prime i}\,\int_0^{R/r}dz\,{z^2\, B(t,z\,r)\over
\sqrt{1+z^2-2z\,\ve{n}^\prime\cdot\ve{n}}}+\OO2\,.
\end{eqnarray}

The computation of these and similar integrals discussed below is straightforward and can be performed by
using
\begin{eqnarray}
\label{expansion}
{1\over\sqrt{1+z^2-2z\,x}}&=&
\left[\,
\begin{array}{l}
\displaystyle{\sum_{n=0}^\infty P_n(x)\,z^n\,,\quad |z|<1\,,}
\\[18pt]
\displaystyle{\sum_{n=0}^\infty P_n(x)\,z^{-n-1}\,,\quad |z|>1\,,}
\end{array}
\right.
\\[10pt]
\label{expansion-for-the-superpotential}
\sqrt{1+z^2-2z\,x}&=&
\left[\,
\begin{array}{l}
\displaystyle{\sum_{n=0}^\infty C^{(-1/2)}_n(x)\,z^n\,,\quad |z|<1\,,}
\\[18pt]
\displaystyle{\sum_{n=0}^\infty C^{(-1/2)}_n(x)\,z^{-n+1}\,,\quad |z|>1\,,}
\end{array}
\right.
\end{eqnarray}
\noindent 
where $P_n(x)$ are Legendre polynomials, $C^{(\alpha)}_n(x)$ are
Gegenbauer polynomials, and
\begin{eqnarray}
\label{integral-nhat-L-P-s}
&&\int_0^{2\pi}d\lambda^\prime\int_0^{\pi}d\theta^\prime\,
\sin\theta^\prime\,\hat n^{\prime L}\,
P_s(\ve{n}^\prime\cdot\ve{n})={4\,\pi\over 2l+1}\,\hat n^L\,\delta^{ls}\,,\quad l\ge0\,,
\\
\label{Gegenbauer}
&&\int_0^{2\pi}d\lambda^\prime\int_0^{\pi}d\theta^\prime\,
\sin\theta^\prime\,
C^{(-1/2)}_s(\ve{n}^\prime\cdot\ve{n})=4\,\pi\,\left(\delta^{0s}+{1\over 3}\,\delta^{2s}\right)\,,\quad s\ge0\,.
\end{eqnarray}
\noindent 
Eq. (\ref{integral-nhat-L-P-s}) can be demonstrated in different
ways: e.g., by using the representation of $\hat n^L$ in terms of spherical
functions and the symmetric trace-free (STF) basis tensors and noting that
$P_s(\ve{n}^\prime\cdot\ve{n})$ can be represented as a sum of associated
Legendre polynomials depending on the spherical coordinates of $\ve{n}$ and
$\ve{n}^\prime$. The orthogonality of the associated Legendre functions can
then be used. Eq.\ (\ref{Gegenbauer}) follows e.g., from the explicit
formulas for the Gegenbauer polynomials $C^{(-1/2)}_s(x)$.

\subsubsection{Internal part}

For an internal point $(t,\ve{x})$ with $r=|\ve{x}|\le R$
the formal solution of (\ref{feq-w})--(\ref{feq-wi}) reads
\begin{eqnarray}
\fl
\label{w-spherical-explicit-internal}
w(t,\ve{x})&=&{4\pi\,G\over r}\,\int_0^rdy\,y^2\,\sigma(t,y)
+4\pi\,G\,\int_r^Rdy\,y\,\sigma(t,y)
\nonumber\\
\fl
&&
+{2\pi\,G\over c^2}\,{\partial^2 \over \partial t^2}\,
\Biggl(
r\,\int_0^rdy\,y^2\,\sigma(t,y)
+{1\over 3r}\,\int_0^rdy\,y^4\,\sigma(t,y)
\nonumber\\
\fl
&&
\phantom{+{2\pi\,G\over c^2}\,{\partial^2 \over \partial t^2}\,\quad}
+\int_r^Rdy\,y^3\,\sigma(t,y)
+{1\over 3}\,r^2\,\int_r^Rdy\,y\,\sigma(t,y)
\Biggr)+\OO4\,,
\\
\fl
\label{wi-spherical-explicit-internal}
w^i(t,\ve{x})&=& {4\pi\,G\over 3}\,\left({x^i\over r^3}\,\int_0^rdy\,y^3\,B(t,y)
+x^i\,\int_r^Rdy\,B(t,y)\right)+\OO2\,.
\end{eqnarray}

\subsubsection{External part} For an external point $(t,\ve{x})$ with $r=|\ve{x}|\ge R$
the formal solution of (\ref{feq-w})--(\ref{feq-wi}) can be simplified so that
\begin{eqnarray}
\fl
\label{w-spherical-explicit}
w(t,\ve{x})&=&{4\pi\,G\over r}\,\int_0^Rdy\,y^2\,\sigma(t,y)
\nonumber\\
\fl
&&
+{2\pi\,G\over c^2}\,{\partial^2 \over \partial t^2}\,
\left(r\,\int_0^Rdy\,y^2\,\sigma(t,y)+{1\over 3r}\,\int_0^Rdy\,y^4\,\sigma(t,y)
\right)+\OO4\,,
\\
\fl
\label{wi-spherical-explicit}
w^i(t,\ve{x})&=& {4\pi\,G\over 3}\,{x^i\over r^3}\,\int_0^Rdy\,y^3\,B(t,y)+\OO2\,.
\end{eqnarray}

\subsubsection{General multiple expansions for the external part}
\label{section-multipole-expansions} As is well known the solution of
(\ref{feq-w})--(\ref{feq-wi}) outside an arbitrary compact matter
distribution admits an expansion in terms of multipole moments (e.g.,
\cite{BlanchetDamour1989}). Such an expansion takes the form
\begin{eqnarray}\label{BD-expansion-w}
w&=&G\sum_{l=0}^\infty {(-1)^l\over l!}
\left[ M_L\,\partial_L {1\over r}+
{1\over 2c^2}\,\ddot M_L
\,\partial_L\,r\right]
+{4\over c^2}\Lambda_{,t}+\OO4,
\end{eqnarray}
\begin{eqnarray}\label{BD-expansion-PPN-Wi}
w^i&=&-G\sum_{l=1}^\infty {(-1)^l\over l!}
\left[
\dot M_{iL-1} \partial_{L-1} {1\over r}+
{l\over l+1} \varepsilon_{ijk} S_{kL-1}
\partial_{jL-1} {1\over r}
\right]
\nonumber \\
&&
-\Lambda_{,i}+\OO2,
\end{eqnarray}
\noindent
where
\begin{eqnarray}\label{Lambda}
\Lambda&=&G \sum_{l=0}^\infty
{(-1)^l\over (l+1)!}\,{2l+1\over 2l+3}
\,{\cal P}_L\,\partial_L {1\over r}.
\end{eqnarray}
\noindent 
The Blanchet-Damour moments, $M_L$ and $S_L$, are given by
\begin{eqnarray}\label{ML-BD}
M_L&=&\int_V \sigma\, \hat x^L d^3x
+{1\over 2(2l+3)} {1\over c^2}\, {d^2\over dt^2}
\int_V \sigma\, \hat x^L\,x^2\,d^3x
\nonumber \\
&&
-{4\,(2l+1)\over (l+1)\,(2l+3)} {1\over c^2}\,
{d\over dt} \int_V \sigma^i \hat x^{iL}\,d^3x,\quad l \ge 0,
\\ \label{SL-BD}
S_L&=&\int_V \varepsilon^{ij\langle a_l} \hat x^{L-1\rangle i}
\,\sigma^j\,d^3s,\quad l \ge 1.
\end{eqnarray}
\noindent 
The additional moments ${\cal P}_L$  are defined by
\begin{eqnarray}\label{PL}
{\cal P}_L=\int_V \sigma^i \, \hat x^{iL}\,d^3x,\quad l\ge0\,.
\end{eqnarray}
\noindent
Here $V$ again denotes the support of the matter distribution.

Since we consider an isolated matter distribution of compact-support it is
well known that according to the local equations of motion the lower
multipole moments satisfy the equations
\cite{DamourSoffelXu1991,KlionerSoffel2000}:
\begin{eqnarray}
\dot M=\OO4\,, \quad \ddot M_i=\OO4\,, \quad \dot S_i=\OO2\,.
\end{eqnarray}
\noindent
It is also clear that $M_i$ can always be chosen to be identically
zero by the choice of the origin of the reference system as 
the post-Newtonian center of mass.

\subsubsection{General skeletonized harmonic gauge}
The terms containing $\Lambda$ can be eliminated from
(\ref{g00})--(\ref{g0i}) by a transformation of the time coordinate
\begin{equation}\label{kill-Lambda}
t^\prime=t-{4\over c^4} \Lambda\,,
\quad
\ve{x}^\prime=\ve{x}\,.
\end{equation}
\noindent
This coordinate transformation obviously retains the harmonics gauge.
This transformation changes the metric tensor as
\begin{eqnarray}
\label{g00'}
g_{00}^\prime&=&g_{00}-{8\over c^4}\,\Lambda_{,t}+\OO5\,,
\\
\label{g0i'}
g_{0i}^\prime&=&g_{0i}-{4\over c^3}\,\Lambda_{,i}+\OO5\,,
\\
\label{gij'}
g_{ij}^\prime&=&g_{ij}+\OO5\,.
\end{eqnarray}
\noindent 
This gauge is called skeletonized harmonic gauge
\cite{DamourSoffelXu1991} in which $\Lambda$-terms do not appear in the
post-Newtonian metric: neither in (\ref{g00}), (\ref{g0i}), nor in the terms
$\OO2$ of (\ref{gij}). In this approximation, the metric is ``skeletonized''
by the Blanchet-Damour moments $M_L$ and $S_L$. However, it is important to
understand that the transformation (\ref{kill-Lambda}) does not change
$g_{ij}$ and, therefore, terms depending on $\Lambda$ are still present in
the terms $\OO4$ in $g_{ij}$.

\subsubsection{Multipole moments for a spherically symmetric matter distribution}
For the spherically symmetric matter distribution
(\ref{sigma})--(\ref{sigma-ij}) one can easily show that
\begin{eqnarray}\label{M-BD}
\fl
M&=&\int_V \sigma\,d^3x
-{1\over 2\,c^2}\,{d^2\over dt^2} N
=4\pi\,\int_0^R dy\,y^2\,\sigma(t,y)-
{2\pi\over c^2}\,{d^2\over dt^2}\,
\int_0^R dy\,y^4\,\sigma(t,y)\,,
\\
\fl
\label{N}
N &=& \int_V \sigma\,r^2\,d^3x
=4\pi\,\int_0^R dy\,y^4\,\sigma(t,y)\,,
\\
\fl
\label{P-spherically-symmetric}
{\cal P}&=&\int_V \sigma^i\,x^i\,d^3x={1\over 2}\,{\dot N}+\OO2
=4\pi\int_0^Rdy\,y^3\,B(t,y)+\OO2\,,
\\
\fl
\label{ML-BD-0}
M_L&=&0,\quad l \ge 1\,,
\\
\fl
\label{SL-BD-0}
S_L&=&0,\quad l \ge 1\,,
\\
\fl
\label{PL-0}
{\cal P}_L&=&0,\quad l \ge 1\,.
\end{eqnarray}
\noindent 
In this case the Blanchet-Damour mass $M$ coincides with the Tolman
mass \cite{KlionerSoffel2000} and thus coincides with the mass
parameter of the Schwarzschild metric as discussed e.g., in
\cite{Weinberg1972}.  The relation ${\cal P}={1\over 2}\,{\dot
  N}+\OO2$ holds for an arbitrary matter distribution and follows from
the Newtonian equation of continuity (see Eq. (\ref{Newtonian-continuity}) below)
and the Ostrogradsky-Gauss theorem.

It is easy to see that $w$ and $w^i$ from
(\ref{w-spherical-explicit})--(\ref{wi-spherical-explicit}) admit
multipole expansions (\ref{BD-expansion-w})--(\ref{Lambda}) with
multipole moments given by (\ref{M-BD})--(\ref{PL-0}).

\bigskip

In the following we work only with the skeletonized harmonic gauge and drop the primes over the coordinates.
Thus, the metric tensor in this gauge at the external point $(t,\ve{x})$ with $|\ve{x}|\ge R$ takes the form
\begin{eqnarray}
\label{g00''-explicit}
g_{00}&=&-1 + {2 \over c^2} {G\,M\over r} - {2 \over c^4} {G^2M^2\over r^2} + \OO5 \, ,
\\
\label{g0i''-explicit}
g_{0i}&=&\OO5\,,
\\
\label{gij''-explicit}
g_{ij}&=&\delta_{ij} \left( 1 + {2 \over c^2} {G\,M\over r} + {2 \over c^4} {G^2M^2\over r^2}\right)
+ {4 \over c^4} \left(q_{ij}+\delta_{ij}\,{G{\ddot N}\over 3r}\right) + \OO5\,.
\end{eqnarray}

\subsection{Computation of $q_{ij}$}

We now come to the computation of $q_{ij}$, as a solution of (\ref{qij}), which
can be split according to
\begin{eqnarray}
\label{qij-split}
q_{ij} &=& q_{ij}^w + q_{ij}^\sigma\,,
\end{eqnarray}
\noindent
where
\begin{eqnarray}
\label{qij-w} \Delta q^w_{ij} &=&  - w_{,i} w_{,j}+ \OO1 \,,
\\
\label{qij-sigma}
\Delta q^\sigma_{ij} &=& - 4 \pi G \sigma^{ij} + \OO1 \,.
\end{eqnarray}

\subsubsection{Computation of $q_{ij}^w$}

The gravitational potential $w$ in the Newtonian approximation is determined by
(\ref{w-spherical-explicit-internal}) where the terms $\OO2$ are omitted.
Since $w = w(t,r)$ we get
\begin{equation}
\label{w-partial-r}
{\partial w\over\partial x^i}=
{x^i\over r}\,{\partial w\over\partial r},\quad
{\partial w\over\partial r}=
\left[\
\begin{array}{l}
\displaystyle{-{GM_r\over r^2}}, \quad r\le R
\\[10pt]
\displaystyle{-{GM\over r^2}},\quad r\ge R
\end{array}
\right.\,,
\end{equation}
\noindent 
where $M_r$  is the mass contained in a sphere of radius $r$
\begin{equation}
M_r=\int_{|\ve{x}|\le r}\sigma\,d^3x={4\pi}\,\int_0^r \sigma(t,y)\,y^2\,dy
\,,
\end{equation}
\noindent
and $M \equiv M_R$ is the total mass of the body.
Note, that $M_r$ is some unknown function of $r$, while $M$ does not depend
on $r$. Therefore, one gets
\begin{eqnarray}
\Delta\,q_{ij}^{w}&=&-w_{,i}w_{,j}=-G^2M^2\, f^2(r)\,{x^ix^j\over r^6}\,,
\end{eqnarray}
\noindent
where
\begin{equation}
f(r) \equiv M_r/M \,,
\end{equation}
\noindent
so that $f(0)=0$ and $f(r)=1$ for $r\ge R$. The solution for $q_{ij}^w$  can be written as
\begin{equation}
\label{qijw-sum}
q_{ij}^w=G^2M^2 \left(I_{ij}+E_{ij}\right),
\end{equation}
\noindent
where $I_{ij}=I_{ij}(t,\ve{x})$ is the potential
with the source defined by the gravitational potential $w(t,r)$ inside the body (for $r\le R$):
\begin{eqnarray}
\fl
I_{ij}&=&{1\over 4\pi}\, \int_{|\ve{x}^\prime|\le R}
f^2(r^\prime)\,{x^{\prime i}\,x^{\prime j}\over r^{\prime 6}}\,
{1\over |\ve{x}-\ve{x}^\prime|}\,d^3x^\prime
\nonumber\\
&=& {1\over 4\pi}\,{1\over
r^2}\,\int_0^{2\pi}d\lambda^\prime\int_0^{\pi}d\theta^\prime\,
\sin\theta^\prime\,n^{\prime i}\,n^{\prime
j}\,\int_0^{R/r}dz\,{f^2(zr)\over
z^2\sqrt{1+z^2-2z\,\ve{n}^\prime\cdot\ve{n}}} \label{I1-ij-integral}
\end{eqnarray}
\noindent 
and $E_{ij}=E_{ij}(t,\ve{x})$ is the potential with the source
defined by the gravitational potential $w(t,r)$ outside the body (for $r\ge
R$):
\begin{eqnarray}
\fl
E_{ij}&=&{1\over 4\pi}\, \int_{|\ve{x}^\prime|\ge R} {x^{\prime
i}\,x^{\prime j}\over r^{\prime 6}}\, {1\over
|\ve{x}-\ve{x}^\prime|}\,d^3x^\prime
\nonumber\\
&=& {1\over 4\pi}\,{1\over
r^2}\,\int_0^{2\pi}d\lambda^\prime\int_0^{\pi}d\theta^\prime\,
\sin\theta^\prime\,n^{\prime i}\,n^{\prime
j}\,\int_{R/r}^{\infty}dz\,{1\over
z^2\sqrt{1+z^2-2z\,\ve{n}^\prime\cdot\ve{n}}}
\nonumber\\
&=& {1\over 4\pi}\,{1\over
r^2}\,\int_0^{2\pi}d\lambda^\prime\int_0^{\pi}d\theta^\prime\,
\sin\theta^\prime\,n^{\prime i}\,n^{\prime
j}\,\int_{0}^{r/R}dz\,{z\over \sqrt{1+z^2-2z\,\ve{n}^\prime\cdot\ve{n}}}\,,
\label{I2-ij-integral}
\end{eqnarray}
\noindent 
where $r^{\prime}=|\ve{x}^\prime|$, $r=|\ve{x}|$,
$\ve{n}^{\prime}=\ve{x}^\prime/r^\prime$, $\ve{n}=\ve{x}/r$.  Both
potentials $I_{ij}$ and $E_{ij}$ are non-zero both inside and outside
of the matter distribution. For each of these two integrals, $I_{ij}(t,\ve{x})$ and $E_{ij}(t,\ve{x})$,
two cases should be considered: the external case with $|\ve{x}|=r\ge R$
(labeled by a superscript '$\exter$') and the internal case for $|\ve{x}|=r\le R$ (labeled by a
superscript '$\inter$'). Straightforward calculations show that
\begin{eqnarray}
\label{Iij-int}
I^\inter_{ij}&=&
{1\over 3\,r^2}\,\delta^{ij}\, \left(r\,\int_0^{r}dy\,{f^2(y)\over
y^2}+r^2\int_{r}^{R}dy\,{f^2(y)\over y^3}\right)
\nonumber\\
&&
+{1\over 5\,r^2}\,\hat{n}^{ij}\, \left({1\over
r}\,\int_0^{r}dy\,f^2(y)+r^4\int_{r}^{R}dy\,{f^2(y)\over y^5}\right) \,,
\\
\label{Iij-ext}
I^\exter_{ij}&=&{1\over 3\,r}\,\delta^{ij}\,\int_0^{R}dy\,{f^2(y)\over y^2} +{1\over
5\,r^3}\,\hat{n}^{ij}\,\int_0^{R}dy\,f^2(y) \,.
\\
\label{Iij-2-int}
\label{Eij-int}
E^\inter_{ij}&=&{1\over 6}\,{1\over R^2}\,\delta^{ij}+{1\over 20}\,{r^2\over
R^4}\,\hat n^{ij}\,,
\\
\label{Iij-2-ext}
\label{Eij-ext}
E^\exter_{ij}&=&{1\over 3}\,{1\over R\,r}\,\delta^{ij}-{1\over 6}\,{1\over
r^2}\,\delta^{ij} +{1\over 4}\,{1\over r^2}\,\hat n^{ij}-{1\over 5}\,{R\over
r^3}\,\hat n^{ij}\,.
\end{eqnarray}
\noindent 
Note, that the integrals in (\ref{Iij-ext}) do not depend on $r$.
Therefore, the dependence of $I^\exter_{ij}$ on $r$ is explicitly found.

\subsubsection{Computation of $q_{ij}^\sigma$}

We now turn to the computation of $q_{ij}^\sigma$ determined by (\ref{qij-sigma}).
Using (\ref{sigma-ij}) we have:
\begin{eqnarray}
\label{qsigmaE-general-external} q_{ij}^{\sigma}&=&G \int_{|\ve{x}^\prime|\le R}
{\sigma^{ij}(t,\ve{x}^\prime) \over |\ve{x}-\ve{x}^\prime|}d^3x^\prime
\nonumber\\
&=&G\,r^2\int_0^{2\pi}d\lambda^\prime\int_0^{\pi}d\theta^\prime\,
\sin\theta^\prime\,\hat{n}^{\prime ij}\, \int_0^{R/r}dz\, {z^2\,A(zr)\over
\sqrt{1+z^2-2z\,\ve{n}^\prime\cdot\ve{n}}}
\nonumber\\
&& -2\,G\,r^2\,\delta^{ij}\int_0^{2\pi}d\lambda^\prime\int_0^{\pi}d\theta^\prime\,
\sin\theta^\prime\, \int_0^{R/r}dz\, {z^2\,C(zr)\over
\sqrt{1+z^2-2z\,\ve{n}^\prime\cdot\ve{n}}}\,.
\end{eqnarray}
\noindent
Here we do not specify explicitly that $A$ and $C$ may also depend on time $t$.
Again two cases $r\ge R$ and $r\le R$ should be considered using
(\ref{expansion})--(\ref{integral-nhat-L-P-s}). For an internal point
with $r\le R$ the integral expression for $q_{ij}^\sigma$ reads
\begin{eqnarray}
\label{qij-sigma-internal}
q_{ij}^{\sigma,\inter}&=&
{4\pi\,G\over5}\,{\hat{n}^{ij}\over r^3}\, \int_0^{r}dy\,y^4\,A(y)
+{4\pi\,G\over5}\,r^2\,\hat{n}^{ij}\,\int_{r}^{R}dy\,{A(y)\over y}
\nonumber\\
&&
-{8\pi\,G\over r}\,\delta^{ij}\,\int_0^{r}dy\, y^2\,C(y)
-8\pi\,G\,\delta^{ij}\, \int_{r}^{R}dy\,y\,C(y)\,.
\end{eqnarray}
For an external point
with $r\ge R$ the integral expression for $q_{ij}^\sigma$ can be simplified to
\begin{eqnarray}
\label{qij-sigma-external} q_{ij}^{\sigma,\exter}&=&
{4\pi\,G\over5}\,{\hat{n}^{ij}\over r^3}\, \int_0^{R}dy\,y^4\,A(y)
-{8\pi\,G\over r}\delta^{ij}\,\int_0^{R}dy\, y^2\,C(y) \,.
\end{eqnarray}
\noindent
Again the integrals on the last line of (\ref{qij-sigma-external}) do
not depend on $r$  and the dependence of $q_{ij}^{\sigma,\exter}$
on $r$ is explicitly given by (\ref{qij-sigma-external}).

For the general spherically symmetric case considered here, $f=f(r)$, $A(r)$
and $C(r)$ are arbitrary functions and no further simplification of the
internal potentials $I^\inter_{ij}$ and $q_{ij}^{\sigma,\inter}$ can be done.
Clearly, the internal potentials $I^\inter_{ij}$, $E^\inter_{ij}$ and
$q_{ij}^{\sigma,\inter}$ are not needed for the derivation of the external
metric. They will be used below for  comparisons in the special case of a
body composed of an incompressible fluid.

\subsection{External metric}

Gathering all the partial results we can now write  the following expression
for the potential $q_{ij}(t,\ve{x})$ at an external point with $|\ve{x}|\ge
R$:
\begin{eqnarray}
\label{qij-external}
q_{ij}^\exter&=&G^2M^2 \left(I_{ij}^\exter+E_{ij}^\exter\right)+q_{ij}^{\sigma,\exter}
\nonumber\\
&=&
{G^2M^2\over 3\,r}\,\delta^{ij}\,\int_0^{R}dy\,{f^2(y)\over y^2} +{G^2M^2\over
5\,r^3}\,\hat{n}^{ij}\,\int_0^{R}dy\,f^2(y)
\nonumber\\
&&
+{G^2M^2\over 3\,R\,r}\,\delta^{ij}-{G^2M^2\over 6\,r^2}\,\delta^{ij}
+{G^2M^2\over 4\,r^2}\,\hat n^{ij}-{G^2M^2R\over 5r^3}\,\hat n^{ij}
\nonumber\\
&&
+{4\pi\,G\over5}\,{\hat{n}^{ij}\over r^3}\, \int_0^{R}dy\,y^4\,A(y)
-{8\pi\,G\over r}\delta^{ij}\,\int_0^{R}dy\, y^2\,C(y)\,.
\end{eqnarray}
\noindent 
All integrals in (\ref{qij-external}) are constants characterizing
the matter distribution under consideration in addition to the mass $M$. Such
additional constants do not appear in usual forms of the external
Schwarzschild metric and either can be eliminated by some coordinate
transformation or vanish in virtue of the local equations of motion.

The dependence of $q_{ij}^\exter=q_{ij}^\exter(t,\ve{x})$ on $\ve{x}$ in
(\ref{qij-external}) is fully explicit. There are terms of the following
type: i) $\delta^{ij}/r$, ii) $\delta^{ij}/r^2$, iii) $\nhat^{ij}/r^2$ and
iv) $\nhat^{ij}/r^3$.  The additional constants appear in terms of types i
and iv. We demonstrate first that the terms of type i cancel with the term in
(\ref{gij''-explicit}) proportional to $\ddot N$ and coming from
$\Lambda_{,t}$ in (\ref{BD-expansion-w}). Collecting all terms of this type
in (\ref{qij-external}) one gets
\begin{eqnarray}
\label{qij-external-1/r}
\left.\phantom{\Biggl|}q_{ij}^\exter\right|_{1/r}&=& {\delta^{ij}\,G\over
3r}\left( -24\,\pi\int_0^{R}dy\, y^2\,C(y) +GM^2\int_0^{R}dy\,{f^2(y)\over
y^2} +{GM^2\over R}\right)
\nonumber\\
&=& -8\pi\,{\delta^{ij}\,G\over 3r} \int_0^{R}dy\,
\left(3y^2\,C(y)-GM\,\sigma(y)\,y\,f(y)\right)
\nonumber\\
&=& -8\pi\,{\delta^{ij}\,G\over 3r} \int_0^{R}dy\,
\left(3y^2\,C(y)+y^3\,\sigma(y)\,{dw(y)\over dy}\right)
\nonumber\\
&=& -8\pi\,{\delta^{ij}\,G\over 3r}\,
{d\over dt}\int_0^{R}dy\,y^3\,B(t,y)
= -{\delta^{ij}\,G\over 3r}\,{\ddot N}\,,
\label{finalres}
\end{eqnarray}
\noindent
where we used the Newtonian local equations of motion
\begin{eqnarray}
\label{Newtonian-continuity}
&&{\partial\over\partial t}\sigma+{\partial\over\partial x^i}\sigma^i= \OO2\,,
\\
&& {\partial\over\partial
t}\sigma^i+{\partial\over\partial x^j}T^{ij}= \sigma\,{\partial\over\partial
x^i}w+\OO2 \,.
\end{eqnarray}
\noindent
The second equation, in the case of spherical symmetry (\ref{sigma})--(\ref{sigma-ij}), can be simplified to
\begin{eqnarray}
\fl
&&{2\over 3}\,{\partial\over\partial r}A(t,r)+{2A(t,r)\over r}
+{\partial\over\partial t} B(t,r)+{\partial\over\partial
r}C(t,r)= \sigma\,{\partial\over\partial r}w(t,r)+\OO2.
\end{eqnarray}
\noindent 
The quantity $N$ appearing in the final result in (\ref{finalres})
is just the moment of inertia of the body defined by (\ref{N}). Comparing
(\ref{finalres}) with  the last term in (\ref{gij''-explicit}) we conclude
that the $1/r$ terms of order $\OO4$ in $g_{ij}$ cancel exactly.

Finally, let us note that the terms of type iv
in (\ref{qij-external}) can be eliminated by a gauge transformation:
\begin{equation}\label{kill-gradients}
t^\prime=t\,,
\quad
{x^\prime}^i=x^i+{1\over c^4}\,\partial_i h\,.
\end{equation}
\noindent
This transformation changes the metric tensor according to
\begin{eqnarray}
\label{g00''}
g_{00}^\prime&=&g_{00}+\OO5\,,
\\
\label{g0i''}
g_{0i}^{\prime}&=&g_{0i}+\OO5\,,
\\
\label{gij''}
g_{ij}^\prime&=&g_{ij}-{2\over c^4} \partial_{ij}h+\OO5\,.
\end{eqnarray}
\noindent 
One can see that the coordinate gauge remains harmonic if the
function $h$ satisfies the condition $\partial_{kk} h=\OO1$. Taking
\begin{eqnarray}
\label{h-r}
h={1\over 30}\,{G\over r}\,\left(
GM^2\,\int_0^{R}dy\,f^2(y)
-GM^2R
+4\pi\,\int_0^{R}dy\,y^4\,A(y)
\right)
\end{eqnarray}
\noindent 
one can eliminate the terms of type iv in (\ref{qij-external}) and in
the metric. The transformation (\ref{kill-gradients}) with $h$ given
by (\ref{h-r}) augments the definition of the skeletonized harmonic
gauge for a spherically symmetric matter distribution in the
post-linear approximation. Note, that both $\Lambda$ appearing in
(\ref{kill-Lambda}) and $h$ in (\ref{kill-gradients}) depend on the
internal structure of the body, while the resulting external metric
does not. Indeed, omitting the primes again, we can see that the
metric tensor at the external point $(t,\ve{x})$ with $|\ve{x}|\ge R$
takes the form
\begin{eqnarray}
\label{g00'-explicit}
g_{00}&=&-1 + {2 \over c^2} {G\,M\over r} - {2 \over c^4} {G^2M^2\over r^2} + \OO5 \, ,
\\
\label{g0i'-explicit}
g_{0i}&=&\OO5\,,
\\
\label{gij'-explicit}
g_{ij}&=&\delta_{ij} \left( 1 + {2 \over c^2} {G\,M\over r} + {1 \over c^4} {G^2M^2\over r^2}\right)
+ {1 \over c^4}\,{G^2M^2\over r^2}\,n^in^j+ \OO5 \, .
\end{eqnarray}
\noindent 
This metric fully agrees with the well-known external Schwarzschild
metric in harmonic coordinates in the corresponding approximation.

\section{The case of a static incompressible fluid}
\label{section-incompressible-metric}

The case of a static body composed of an incompressible fluid is often
discussed in the literature when dealing with the internal Schwarzschild
solution \cite{Weinberg1972}. It is well known that for a static
incompressible fluid the four functions describing the matter distribution in
(\ref{sigma})--(\ref{sigma-ij}) are time-independent and read
\begin{eqnarray}
\fl \label{sigma-incompressible} \sigma(r)&=&\kappa\,\left(1+{1\over
c^2}\left(2w+3p\right)\right)+\OO4= \kappa\,\left(1+{1\over 2c^2}\,
{GM\over R}\left(9-5\eta^2\right)\right)+\OO4\,,
\\
\fl
\label{A-incompressible}
A(r)&=&\OO2\,,
\\
\fl
\label{B-incompressible}
B(r)&=&\OO2\,,
\\
\fl \label{C-incompressible} C(r)&=&p+\OO2={1\over 2}\,{GM\over
R}\,\left(1-\eta^2\right)\,\kappa+\OO2\,,
\end{eqnarray}
\noindent 
where $\eta\equiv r/R$, $\kappa={\rm const}$ is the invariant
density (rest mass plus internal energy density), $p=p(r)$ is the isotropic
pressure that can be computed from the condition of hydrostatic equilibrium
${dp/dr}=\kappa\,{dw/dr}+\OO2$ with the boundary condition $p(R)=0$. The
well-known Newtonian formula for the internal potential,
$w={1\over 2}\,{GM\over R}\,\left(3-\eta^2\right)+\OO2$, was used here and in
(\ref{sigma-incompressible}).

The equations that define the gravitational potentials simplify for the
static incompressible fluid case to
\begin{eqnarray}
\label{w-incompressible}
\fl
w(t,\ve{x})&=&
\left[
\begin{array}{l}
\displaystyle{ {1\over 2}\,{GM\over R}\,\left(3-\eta^2\right) +{3\over
8c^2}\,{G^2M^2\over R^2}\,\left(1-\eta^2\right)^2 +\OO4 \,,\quad r\le R\,, }
\\[12pt]
\displaystyle{
{GM\over r}+\OO4\,,\quad r\ge R\,,
}
\end{array}
\right.
\\[10pt]
\fl
\label{wi-incompressible}
w^i(t,\ve{x})&=&\OO2\,,
\\[10pt]
\fl
\label{qij-incompressible}
q^{ij}(t,\ve{x})&=&
G^2M^2
\nonumber\\
\fl
&&
\times
\left[
\begin{array}{l}
\displaystyle{
{1\over R^2}\,\eta^2\left({3\over 20}-{1\over 14}\,\eta^2\right)\,\hat n^{ij}
-{1\over 2 R^2}\left(1-\eta^2+{1\over 3}\,\eta^4\right)\,\delta^{ij}
+\OO1\,,
\ r\le R\,,
}
\\[12pt]
\displaystyle{ {1\over 4r^2}\left(\hat n^{ij}-{2\over 3}\,\delta^{ij}\right)
-{2\over 35}\,R\,\partial_{ij}\left({1\over r}\right)+\OO1\,,\ r\ge R\,, }
\end{array}
\right.
\end{eqnarray}
\noindent 
where the mass $M$ is defined by
\begin{eqnarray}\label{mass}
\fl M&=&4\pi\,\int_0^Rdy\,y^2\,\sigma(t,y)+\OO4 ={4\over
3}\,\pi\,R^3\,\kappa\left(1+{3GM\over c^2R}\right)+\OO4\,.
\end{eqnarray}
\noindent 
Here we used the fact that for static incompressible fluid
$f(r)=\eta^3+\OO1$.

It is important to see that for a static incompressible fluid ${\cal P} =0$
and therefore $\Lambda=0$ (see Eqs.~(\ref{P-spherically-symmetric}) and
(\ref{Lambda})\,). It means that no additional time transformation
(\ref{kill-Lambda}) is needed to bring the external metric in the usual form
of the Schwarzschild solution in harmonic coordinates. In the gauge
transformation of spatial coordinates (\ref{kill-gradients})-(\ref{h-r}), one
should take $h=-{G^2M^2R\over 35\,r}$. This eliminates the last term in
(\ref{qij-incompressible}) for $q^{ij}$ for an external point. In this way
the metric outside of the body again coincides with
(\ref{g00'-explicit})--(\ref{gij'-explicit}) and agrees with the well-known
Schwarzschild solution.

\medskip
For a point inside the body with $r\le R$,\ Eqs.\
(\ref{w-incompressible})--(\ref{qij-incompressible}) together with the
definitions (\ref{g00})--(\ref{gij}) allow us to write
\begin{eqnarray}
\fl
\label{g00-incompressible}
g_{00}&=&-1 + {1 \over c^2}\,{G\,M\over R}\,(3-\eta^2) - {1 \over 4c^4}\,{G^2M^2\over R^2}\,\left(15-6\eta^2-\eta^4\right)+ \OO5 \, ,
\\
\fl
\label{g0i-incompressible}
g_{0i}&=&\OO5\,,
\\
\fl
\label{gij-incompressible}
g_{ij}&=&\delta_{ij} \left( 1 +
{1 \over c^2}\,{G\,M\over R}\,(3-\eta^2)
+{1 \over 12c^4}\,{G^2M^2\over R^2}\,\left(39-30\eta^2+7\eta^4\right)
\right)
\nonumber\\
\fl
&&
+{1 \over 35c^4}\,{G^2M^2\over R^2}\,\eta^2\,\left(21-10\eta^2\right)\,\hat n^{ij} + \OO5 \, .
\end{eqnarray}
\noindent 
As expected we see that the internal metric depends on the radius
of the body $R$ as it is the case also in the Newtonian limit.

\section{Derivation of the metric from the exact solution in the case of a static incompressible fluid}
\label{section-incompressible}

\def\cA{{\cal A}}
\def\cB{{\cal B}}
\def\cD{{\cal D}}
\def\cN{{\cal N}}

It is well known that both internal and external Schwarzschild solutions
for the case of the static incompressible fluid can be written as exact
solutions.  In this Section we will compare our results 
(\ref{w-incompressible})--(\ref{gij-incompressible}) for the static 
incompressible fluid case with those that can be
found in the literature (e.g., \cite{Weinberg1972} where
standard Schwarzschild coordinates are used). 

\subsection{Metric tensor in standard coordinates}

For this Schwarzschild problem the metric tensor in standard coordinates
$(t,\rho,\vartheta,\lambda)$ is of the form
\begin{equation}
\label{schwarzschild-standard}
ds^2 = - \cB(\rho) c^2 dt^2 + \cA(\rho) d\rho^2 + \rho^2(d\vartheta^2 + \sin^2 \vartheta
d\lambda^2) \, .
\end{equation}
\noindent
Let the radius of the body be at $\rho = a$. Then the internal metric
for $\rho\le a$ is given by (e.g., see Sections 8.2 and 11.9 of \cite{Weinberg1972}):
\begin{eqnarray}
\label{cA-}
\cA^-(\rho) &=& \left( 1 - {2m \rho^2 \over a^3 }\right)^{-1}\,, \\
\label{cB-}
\cB^-(\rho) &=& {1 \over 4}
\left[ 3 \left( 1 - {2m \over a} \right)^{1/2} - \left( 1 - {2 m \rho^2
\over a^3} \right)^{1/2} \right]^2 \,,
\end{eqnarray}
\noindent
and the external metric for $\rho\ge a$ reads
\begin{eqnarray}
\label{cA+}
\cA^+(\rho)&=&{\left(1 - {2m \over \rho}\right)}^{-1}\,,
\\
\label{cB+}
\cB^+(\rho) &=& 1 - {2m \over \rho} \, .
\end{eqnarray}
\noindent
Here $m = GM/c^2$ and
\begin{equation}\label{mass-standard}
M = 4 \pi \int^a_0 \kappa \rho^2 d\rho  = {4 \pi \over 3} \kappa a^3 \, .
\end{equation}
\noindent
Below we will show that this expression for $M$ is in accordance with
Eq. (\ref{mass}) above.

 \subsection{Transformation to harmonic coordinates}

Our goal now is to transform this solution into harmonic coordinates. It is well known that the
transformation between standard and harmonic coordinates only affects the radial coordinate $r=r(\rho)$.
The transformation of the radial coordinate brings the metric (\ref{schwarzschild-standard}) into the form:
\begin{eqnarray}
\label{g00-harm}
g_{00} &=& - \cB\, , \\
\label{g0i-harm}
g_{0i} &=& 0\, ,\\
\label{gij-harm}
g_{ij} &=& \cD \, \delta_{ij} +  \cN\, n^i n^j \, ,
\end{eqnarray}
\noindent
where
\begin{equation}
\cD = {\rho^2 \over r^2} \, , 
\qquad 
\label{N-potential}
\cN = \left[ {\left({dr\over d\rho}\right)}^{-2}\,\cA  - {\rho^2 \over r^2} \right] \, .
 \end{equation}

The transformation from the standard radial coordinate $\rho$ to some
harmonic coordinate $r = r(\rho)$ is determined by the second-order
differential equation (e.g., Section 8.1 of \cite{Weinberg1972}):
\begin{equation}\label{trans}
{d \over d\rho} \left( \rho^2\, \cB^{1/2}\, \cA^{-1/2}\, {dr \over d\rho} \right) = 2\,
\cA^{1/2}\, \cB^{1/2}\, r \, .
\end{equation}
\noindent
Clearly this gives two distinct differential equations for $\rho\le a$
and $\rho\ge a$ according to (\ref{cA-})--(\ref{cB-}) and
(\ref{cA+})--(\ref{cB+}).  We will now determine the function
$r(\rho)$ satisfying these two equations such that both $r(\rho)$ and its
derivative $dr/d\rho$ are continuous at the stellar surface at
$\rho = a$ or $r(a) = R$. According to (\ref{trans}) this is needed to
keep the metric in harmonic coordinates continuous at $r(a)=R$. 
In the following we will consistently
neglect all terms proportional to $m^3$ (or equivalently $c^{-6}$).

For both internal and external solutions we start with the ansatz
\begin{equation}
\label{r-rho-ansatz}
r=\rho\left(1+m\,b(\rho)+m^2\,c(\rho)\right)+{\cal O}(m^3)\,,
\end{equation}
\noindent
where $b(\rho)$ and $c(\rho)$ are unknown functions to be determined
from the differential equation (\ref{trans}) and boundary conditions.

For the external case we substitute the external metric
(\ref{cA+})--(\ref{cB+}) and the ansatz (\ref{r-rho-ansatz}) into
(\ref{trans}), expand in powers of $m$, neglect terms ${\cal O}(m^3)$
and get the following general solutions of the resulting second-order
differential equations for $b(\rho)$ and $c(\rho)$:
\begin{eqnarray}
b^+(\rho)&=&
C_2^+
-{1\over\rho}
-{C_1^+\over 3\rho^3}\,,
\\
c^+(\rho)&=&
C_4^+
-{C_2^+\over\rho}
-{C_3^+\over 3\rho^3}
-{2C_1^+\over 3\rho^4}\,,
\end{eqnarray}
\noindent
where $C_i^+$ are four arbitrary constants. A similar procedure for the internal metric 
(\ref{cA-})--(\ref{cB-}) gives
\begin{eqnarray}
b^-(\rho)&=&
{\rho^2\over 2\,a^3}
+C_2^-
-{C_1^-\over 3\rho^3}\,,
\\
c^-(\rho)&=&
{15\over 28}\,{\rho^4\over a^6}
-{3\over 20}\,{\rho^2\over a^4}
+{C_2^-\over 2}\,{\rho^2\over a^3}
+C_4^-
+{C_1^-\over 3\,a^3\,\rho}
-{C_3^-\over 3\rho^3}\,,
\end{eqnarray}
\noindent
where $C_i^-$ are four arbitrary constants (generally speaking different from $C_i^+$). Note that
both $C_i^-$ and $C_i^+$ are not dimensionless. Any values of $C_i^-$ and $C_i^+$ can be taken to 
satisfy the differential equation (\ref{trans}). However, all these constants can be fixed from four
boundary conditions:
\begin{itemize}
\item[1.] $r(\rho)$ is equal to $\rho$ at spatial infinity:
$\lim\limits_{\rho\to\infty} r(\rho)=\rho$ or, equivalently $\lim\limits_{\rho\to\infty} b^+(\rho)=0$ and $\lim\limits_{\rho\to\infty} c^+(\rho)=0$; 
\item[2.] $r(\rho)$ is regular for $\rho=0$: $r(0)=0$ or, equivalently, $\lim\limits_{\rho\to0} \rho\,b^-(\rho)=0$ and $\lim\limits_{\rho\to0} \rho\,c^-(\rho)=0$; 
\item[3.] $r(\rho)$ is continuous at $\rho=a$: $b^-(a)=b^+(a)$ and $c^-(a)=c^+(a)$; 
\item[4.] the derivative of $r(\rho)$ is continuous at $\rho=a$: 
$\left.{db^-(\rho)\over d\rho}\right|_{\rho=a}=\left.{db^+(\rho)\over d\rho}\right|_{\rho=a}$ and 
$\left.{dc^-(\rho)\over d\rho}\right|_{\rho=a}=\left.{dc^+(\rho)\over d\rho}\right|_{\rho=a}$. 
\end{itemize}
\noindent
With these boundary conditions we have:
\begin{eqnarray}
\label{r-}
r^- &=& \rho \left( 1 - {3 \over 2} {m \over a} + {1 \over 2} {m \rho^2 \over
a^3} + {1 \over 4} {m^2 \over a^2} - {9 \over 10} {m^2 \rho^2 \over a^4} +
{15 \over 28} {m^2 \rho^4 \over a^6 }\right)  + {\cal O}(m^3) \, , 
\\
\label{r+}
r^+ &=& \rho\left(1  - {m\over \rho}  -  {4 \over 35} {a
\over \rho}{m^2 \over \rho^2} \right) + {\cal O}(m^3) \, .
\end{eqnarray}
\noindent
This also determines the relation between $a$
(the stellar radius in $\rho$) and $R$ (the stellar radius in $r$):
\begin{eqnarray}
R &=& a - m - {4\over35}\,{m^2\over a} + {\cal O}(m^3) \, , 
\nonumber \\
a &=& R + m + {4\over35}\,{m^2\over R} + {\cal O}(m^3) \, .
\end{eqnarray}
\noindent
From this we see that both definitions for the
mass $M$, (\ref{mass}) and  (\ref{mass-standard}) are in accordance with each
other.

Note that the first derivative of ${\cal A}$ is not continuous at $\rho=a$:
$\left.{d\cA^+\over d\rho}\right|_{\rho=a}\neq
\left.{d\cA^-\over d\rho}\right|_{\rho=a}$. Interestingly, this is compensated by
a discontinuity of the second derivative of $r(\rho)$ at $\rho=a$ so
that the resulting harmonic metric and its first derivatives are
continuous.

Here we have only worked in an approximation neglecting terms ${\cal O}(m^3)$. Let us note that
the differential equation (\ref{trans}) outside the star has the solution (e.g., \cite{BicakKatz2005})
\begin{eqnarray}
&&r = C_1 (\rho - m) + C_2 F(\rho)\,,\quad C_i={\rm const}\,,
\\ 
&&F(\rho) \equiv
 \left[ (\rho - m) \ln (1 - 2m/\rho) + 2m
\right] = - m \sum_{i=2}^\infty {2^i (i-1) \over i (i+1)} (m/\rho)^i \,.
\end{eqnarray}
\noindent
Inside the star (\ref{trans}) can be transformed into a Heun equation for which the
solutions are also known (see also \cite{Hernandez-PastoraMartinRuiz2002}).

Using the coordinate transformations (\ref{r-})--(\ref{r+}) and the metric tensor
(\ref{g00-harm})--(\ref{N-potential}) we can now derive the explicit expressions for the metric tensor in 
harmonics coordinates.

\subsection{Internal metric}

With these results the internal metric is given by
\begin{equation}
\fl
g_{00}^{-} = - \cB^{-} = -1 + {m \over R} (3 - \eta^2) - {m^2 \over 4 R^2} (15 -
6 \eta^2 - \eta^4)+ {\cal O}(m^3) \, ,
\end{equation}
\noindent
from which we derive (\ref{w-incompressible}) for $r \le R$ in
virtue of $w/c^2 = - {1 \over 2}\ln(-g_{00})+ {\cal O}(m^3)$.
Here again $\eta\equiv r/R$. Furthermore,
\begin{eqnarray}
\fl
\cD^{-}(r) &=&  1 + {m \over R}\,(3 - \eta^2) 
+{m^2 \over 2R^2}\,\left({13\over 2}-{27\over 5}\,\eta^2+{19\over 14}\,\eta^4\right)
+ {\cal O}(m^3)\,\\
\fl
\cN^{-}(r) &=& {1 \over 35} {m^2  \over R^2} \, \eta^2 (21 - 10 \eta^2) + {\cal O}(m^3) \, .
\end{eqnarray}
\noindent
These equations and (\ref{gij-harm}) allows one to recover our previous result 
(\ref{gij-incompressible}).

\subsection{External metric}

The metric component $g_{00}^+=-\cB^+$ coincides with (\ref{g00''-explicit}) and
\begin{eqnarray}
\fl
\cD^+(r)  &=& 1 + {2m \over r} + {m^2 \over r^2}\,\left(1+{8 \over 35} {R \over r}
\right) + {\cal O}(m^3)\,, 
\label{D+}
\\
\fl
\cN^+(r) &=& {m^2 \over r^2}\,\left( 1 - {24\over 35}\, {R \over r} \right) +
{\cal O}(m^3) \,.
\label{N+}
\end{eqnarray}
\noindent
These equations and (\ref{gij-harm}) agree with our previous result (\ref{gij'-explicit}).
It is easy to see that the metric in harmonics coordinates and its first derivative is continuous at $r=R$.

\subsection{Computation of $q_{ij}$}

It is interesting to check if we can also recover our expression (\ref{qij-incompressible}) for $q_{ij}$
for a static incompressible fluid. From the definition of $q_{ij}$, Eq.\ (\ref{gij}), one obtains
\begin{equation}
q_{ij} = {c^4 \over 4} \left[ \left( \cD - 1 - {2w \over c^2} - {2w^2 \over
c^4} \right) \delta_{ij} + \cN\, {x^i x^j \over r^2} \right]+ {\cal O}(m^3) \, .
\end{equation}
\noindent
which immediately gives (\ref{qij-incompressible}).

\section{Conclusions}
\label{section-conclusions}

We have treated the gravitational field of some spherically symmetric matter
distribution in harmonic coordinates to post-linear order. We started with
the general case in which the matter distribution might be time dependent and
left the form of the energy-momentum tensor open. The metric tensor was
derived explicitly for both the interior and the exterior region. In the
exterior region the metric tensor can be expanded in terms of the
Blanchet-Damour mass and it is demonstrated explicitly how the usual external
Schwarzschild field can be derived from the field equations. Terms depending
on the internal structure appear in several places in intermediate calculations
and it was shown how they can be removed with additional gauge
transformations or how such terms cancel exactly in virtue of the local
equations of motion.

The results of this paper should be considered as an intermediate step in the
derivation of the post-linear metric (\ref{g00})--(\ref{gij}) for a body
possessing full multipole structure, i.e., having arbitrary mass and spin
moments, $M_L$ and $S_L$. This would be a generalization of the well-known
post-Newtonian multipole expansions of Blanchet and Damour
\cite{BlanchetDamour1989} (discussed also in Section
\ref{section-multipole-expansions}) and the post-Minkowskian ones derived by
Damour and Iyer \cite{DamourIyer1991}. Such a metric is e.g., required for
relativistic modeling of future space astrometric projects aiming at
nanoarcsecond accuracies.

\begin{acknowledgments}
This work was partially supported by the BMWi grants 50\,QG\,0901 and
50\,QG\,1402 awarded by the Deutsche Zentrum f\"ur Luft- und Raumfahrt
e.V. (DLR) as well as by a grant of the Deutsche Forschungsgemeinschaft
(DFG).
\end{acknowledgments}


\end{document}